\documentclass[11pt,a4paper]{article}
\usepackage{amsmath,amsthm,amssymb,amsfonts,mathrsfs,eucal,amsthm}
\usepackage{dsfont}
\usepackage{graphicx,hyperref}
\usepackage[stable]{footmisc}

\def\hrho{\widehat{\rho}}
\def\hiI{\widehat{{\mathscr{I}}}}

\def\states{\mathfrak{S}}

\def\Tr{\operatorname{Tr}}
\def\d{\operatorname{d}}
\def\>{\rangle}
\def\<{\langle}
\def\sH{\mathscr{H}}

\def\geq{\geqslant}
\def\leq{\leqslant}
\def\be{\begin{equation}}
\def\ee{\end{equation}}
\def\bea{\begin{eqnarray}}
\def\eea{\end{eqnarray}}
\def\ben{\begin{eqnarray*}}
\def\een{\end{eqnarray*}}
\def\mE{\mathcal{E}}
\def\mD{\mathcal{D}}
\def\rank{\operatorname{rank}}

\def\N#1{\left|\!\left|{#1}\right|\!\right|}
\def\supp{\mathsf{supp}}
\def\id{\operatorname{id}}
\def\iI{\mathscr{I}}
\def\tr{\operatorname{Tr}} \def\d{\operatorname{d}}
\def\X{\mathcal{X}}
\def\eps{\varepsilon}
\def\ra{\rightarrow}

\def\lra{\leftrightarrow}
\def\I{\widetilde{I}}
\def\F{\mathsf{F}_D^\ra}

\def\P{\mathfrak{p}}

\def\reff#1{(\ref{#1})}

\def\B{\mathfrak{b}}

\def\openone{\mathds{1}}

\renewcommand{\qedsymbol}{\nobreak \ifvmode \relax \else
      \ifdim\lastskip<1.5em \hskip-\lastskip
      \hskip1.5em plus0em minus0.5em \fi \nobreak
      \vrule height0.75em width0.5em depth0.25em\fi}

\newtheorem{theorem}{Theorem}
\newtheorem{corollary}{Corollary}
\newtheorem{lemma}{Lemma}

\theoremstyle{remark}

\theoremstyle{definition}
\newtheorem{definition}{Definition}

\begin{document} 
\title{Distilling entanglement from arbitrary
  resources}  
\author{Francesco Buscemi\footnote{Institute for
    Advanced Research, University of Nagoya, Chikusa-ku, Nagoya
    464-8601, Japan (e-mail:buscemi@iar.nagoya-u.ac.jp)} \and
  Nilanjana Datta\footnote{Statistical Laboratory, University of
    Cambridge, Wilberforce Road, Cambridge CB3 0WB, UK
    (e-mail:n.datta@statslab.cam.ac.uk)}}

\date{\today} 

\maketitle

\begin{abstract} We obtain the general formula for the optimal rate at
  which singlets can be distilled from any given noisy and arbitrarily
  correlated entanglement resource, by means of local operations and
  classical communication (LOCC).  Our formula, obtained by employing
  the quantum information spectrum method, reduces to that derived by
  Devetak and Winter, in the special case of an independent and
  identically distributed resource. The proofs rely on a one-shot
  version of the so-called ``hashing bound'', which in turn provides
  bounds on the one-shot distillable entanglement under general LOCC.
\end{abstract}


\section{Introduction}

A fundamental problem in entanglement theory is to determine how to
optimally convert entanglement, shared between two distant parties,
Alice and Bob, from one form to another. Entanglement manipulation is
the process by which Alice and Bob convert an initial bipartite state
$\rho_{AB}$ which they share, to a required target state
$\sigma_{AB}$, using local operations and classical communication
(LOCC). If the target state $\sigma_{AB}$ is a maximally entangled
state, then the protocol is called entanglement distillation, whereas
if the initial state $\rho_{AB}$ is a maximally entangled state, then
the protocol is called entanglement dilution. Optimal rates of these
protocols, referred to as the {\em{distillable entanglement}} and
{\em{entanglement cost}} of the state $\rho_{AB}$, respectively, were
originally evaluated under $(i)$ the assumption that the entanglement
resource accessible to Alice and Bob was independent and identically
distributed, that is it consisted of multiple, independent and
identical copies, i.e., tensor products $\rho_{AB}^{\otimes n}$, of
the initial bipartite state, and under $(ii)$ the requirement that the
final state of the protocol is equal to $n$ copies of the desired
target state $\sigma_{AB}^{\otimes n}$ with asymptotically vanishing
error in the limit $n \rightarrow \infty$. The distillable
entanglement and entanglement cost computed in this manner are two
asymptotic measures of entanglement of the state
$\rho_{AB}$. Moreover, in the case in which $\rho_{AB}$ is pure, these
two measures of entanglement coincide and are equal to the von Neumann
entropy of the reduced state on any one of the subsystems, $A$ or $B$.

The practical ability to transform entanglement from one form to
another is useful for many applications in quantum information
theory. However, it is not always justified to assume that the
entanglement resource available consists of states which are multiple
copies (and hence tensor products) of a given entangled state. More
generally an entanglement resource is characterized by an arbitrary
sequence of bipartite states which are not necessarily of the tensor
product form.  Sequences of bipartite states on $AB$ are considered to
exist on Hilbert spaces $\sH_A^{\otimes n} \otimes\sH_B^{\otimes n}$
for $n \in \{1,2,3 \ldots\}$. The asymptotic entanglement cost of such
an arbitrary sequence of pure bipartite states was evaluated in
\cite{gbnd1}, whereas the corresponding cost for the more general case
in which the states in the sequence are allowed to be mixed, was
evaluated in \cite{gbnd2}.  As regards entanglement distillation,
Hayashi \cite{haya_conc} evaluated the optimal rate of entanglement
distillation for an arbitrary sequence of pure states. Moreover,
Matsumoto~\cite{matsu} considered the case in which the input mixed
states are supported on the symmetric subspace. This was further
generalized by Brandao and Eisert, who considered the case of
permutationally invariant mixed states~\cite{brandeis}.

In this paper we evaluate the asymptotic distillable entanglement for
a sequence of arbitrary states in two different scenarios: $(i)$ under
one-way (or forward) LOCC, that is, when classical information can
only be sent from Alice to Bob, and $(ii)$ under two-way (or general)
LOCC, that is when both Alice and Bob can send classical information
to each other (possibly multiple times). The resulting expressions for
the distillable entanglement consitute the main results of this paper.
A useful tool for the study of entanglement manipulation in this
general scenario is provided by the Information Spectrum
method~\cite{info-spec}. This method was introduced in Classical
Information Theory by Verdu and Han and has been extended to quantum
information theory first by Ogawa, Nagaoka, and Hayashi. The power of
the information spectrum approach comes from the fact that it does not
depend on the specific structure of sources, channels or entanglement
resources employed in information theoretical protocols.  An important
step on the way to our main result, is to obtain bounds on the
distillable entanglement in the ``one-shot'' scenario, in which Alice
and Bob aim to convert a {\em{single}} copy of a desired target state
$\rho_{AB}$ which they share, to a maximally entangled state, using
LOCC. The logarithm of the maximum rank of the maximally entangled
state which can be thus obtained with a fixed, finite accuracy, is
defined as the one-shot distillable entanglement of $\rho_{AB}$. Our
first result in this context is the one-shot analogue of the
well-known Hashing bound \cite{devetak-winter}, which provides a lower
bound on the distillable entanglement under one-way LOCC. Further, we
obtain more stringent lower bounds on the one-shot distillable
entanglement (both under one-way and two-way LOCC) by first allowing
the initial state to be pre-processed by means of a suitable LOCC map
and then distilling entanglement from the resultant state.  We also
obtain upper bounds to the one-shot distillable entanglement both
under one-way and two-way LOCC. For the case of an arbitrary sequence
of bipartite states, the lower and upper bounds, obtained in the
one-shot scenario, independently converge to the expression for the
distillable entanglement, in the asymptotic limit. Finally, we can
retrieve the well-known expression of the distillable entanglement for
an i.i.d. resource, given in terms of the regularized coherent
information, from our main result by an application of the Generalized
Stein's Lemma \cite{brandao-plenio}.  The paper is organized as
follows. In Section \ref{prelim} we introduce the necessary
definitions and notations. In Section \ref{section_main} we discuss
the protocol of entanglement distillation, and in Section
\ref{one_bounds} we obtain bounds on the one-shot distillable
entanglement. Finally, in Section \ref{asymp} we state and prove our
main results and show that these reduce to the known results in the
i.i.d. scenario. Appendices A and B contain some detailed derivations.

\section{Definitions and notations}
\label{prelim}

\subsection{Mathematical preliminaries}

Let ${\cal B}(\sH)$ denote the algebra of linear operators acting on a
finite--dimensional Hilbert space $\sH$ and let $\states(\sH)$ denote
the set of positive operators of unit trace (states) acting on $\sH$.
Throughout this paper we restrict our considerations to
finite-dimensional Hilbert spaces, and we take the logarithm to base
$2$.

For given orthonormal bases $\{|i^A\rangle\}_{i=1}^d$ and
$\{|i^B\rangle\}_{i=1}^d$ in isomorphic Hilbert spaces
$\sH_A\simeq\sH_B\simeq\sH$ of dimension $d$, we define a
maximally entangled state (MES) of rank $M \le d$ to be
\begin{equation}\label{MES-M}
|\Psi_M^{AB}\>= \frac{1}{\sqrt{M}} \sum_{i=1}^M |i^A\rangle\otimes |i^B\rangle.
\end{equation}
When $M=d$, for any given operator $O\in\mathcal{B(\sH)}$, the
following relation can be shown by direct inspection:
\begin{equation}\label{ricochet}
  (O\otimes\openone)|\Psi_d^{AB}\>=(\openone\otimes O^T)|\Psi_d^{AB}\>,
\end{equation}
where $\openone$ denotes the identity operator, and $O^T$ denotes the
transposition with respect to the basis fixed by
eq.~\reff{MES-M}. Moreover, for any given pure state $|\phi\>$, we
denote the projector $|\phi\>\<\phi|$ simply as $\phi$.

The trace distance between two operators $A$ and $B$ is given by
\begin{equation}\nonumber
  \N{A-B}_1 := \tr\bigl[\{A \ge B\}(A-B)\bigr] - \tr\bigl[\{A <
  B\}(A-B)\bigr],
\end{equation}
where $\{A\ge B\}$ denotes the projector on the
subspace where the operator $(A-B)$ is non-negative, and
$\{A<B\}:=\openone-\{A\ge B\}$. The fidelity of two states $\rho$ and
$\sigma$ is defined as
\begin{equation}\label{fidelity-aaa}
F(\rho, \sigma):= \tr \sqrt{\sqrt{\rho} \sigma \sqrt{\rho}}
=\N{\sqrt{\rho}\sqrt{\sigma}}_1.
\end{equation}
The trace distance between two states $\rho$ and $\sigma$ is related
to the fidelity $F(\rho, \sigma)$ as follows (see
e.~g.~\cite{nielsen}):
\begin{equation}
  1-F(\rho,\sigma) \leq \frac{1}{2} \N{\rho -
    \sigma}_1 \leq \sqrt{1-F^2(\rho, \sigma)},
\label{fidelity}
\end{equation}
where we use the notation $F^2(\rho, \sigma) = \bigl(F(\rho,\sigma)
\bigr)^2$. 

\subsection{Relative entropies and coherent
  information}\label{entropies}
 Our results on the distillable entanglement are expressed in terms
of the following entropic quantities.  For any $\rho,\sigma\ge 0$,
any $0\le P\le\openone$, and any $\alpha\in(0,\infty)\backslash\{1\}$,
we define the following entropic function (related to the
quasi-entropies introduced by Petz in~\cite{petz})
\begin{equation}\label{quasi-ent}
S_\alpha^P(\rho\|\sigma):=\frac{1}{\alpha-1}\log\Tr[\sqrt{P}\rho^\alpha\sqrt{P}\sigma^{1-\alpha}].
\end{equation}
Notice that for $P=\openone$, the function defined above reduces to the well-known R\'enyi relative entropy of order $\alpha$.

In this paper, in particular,
\begin{equation}\label{eq:asda}
S_0^P(\rho\|\sigma):=\lim_{\alpha\searrow 0}S_\alpha^P(\rho\|\sigma),
\end{equation}
plays an important role. Note that
\begin{equation}
S_0^P(\rho\|\sigma)=-\log\Tr[\sqrt{P}\Pi_\rho\sqrt{P}\ \sigma],
\end{equation}
where $\Pi_\rho$ denotes the projector onto the support of $\rho$. 
Further,
\be
S_0^{\openone}(\rho\|\sigma) = S_0(\rho\|\sigma) := - \log (\tr \Pi_\rho \sigma),
\ee
which is the relative R\'enyi entropy of order zero.

In the following we obtain bounds on the distillable entanglement in
terms of two ``smoothed'' quantities, which are derived
from~\reff{eq:asda}, for any $\delta\ge 0$, as
\begin{equation}\label{eq:i}
  I^{A\to B}_{0,\delta}(\rho^{AB}):=\max_{\bar\rho^{AB}\in \B(\rho^{AB};\delta)}\min_{\sigma^B\in\states(\sH_B)}S_0(\bar\rho^{AB}\|\openone_A\otimes\sigma^B),
\end{equation}
and
\begin{equation}\label{eq:itilda}
  \I_{0,\delta}^{A\to B}(\rho^{AB}):=\max_{P\in \P(\rho^{AB};\delta)}\min_{\sigma^B\in\states(\sH_B)}S_0^P(\rho^{AB}\|\openone_A\otimes\sigma^B),
\end{equation}
where
\begin{equation}
 \B(\rho;\delta):=\{\sigma:\sigma\ge 0,\ \Tr[\sigma]\le 1,\ F^2(\rho,\sigma)\ge1-\delta^2\},\label{ball}
\end{equation}
and
\begin{equation}
  \P(\rho;\delta):=\{P:0\le P\le\openone,\ \Tr[P\rho]\ge1-\delta\}.\label{P-ball}
\end{equation}
Note that, in~\reff{ball}, the definition of fidelity~\reff{fidelity-aaa} has been naturally extended to subnormalized density operators.
Such smoothed quantities are needed in order to allow for a finite accuracy (i.e. non-zero error) in the protocol, which is a natural requirement in the one-shot regime. 

For any given $\delta >0$, we refer to $I^{A\to
  B}_{0,\delta}(\rho^{AB})$ and $\I_{0,\delta}^{A\to B}(\rho^{AB})$ as
smoothed zero-coherent informations. These nomenclatures are justified
by anology with the coherent information as follows. For $\delta =0$,
both the above quantities reduce to
\begin{equation}\label{zero-coh}
  I^{A\to B}_{0}(\rho^{AB}):=\min_{\sigma^B\in\states(\sH_B)}S_0(\rho^{AB}\|\openone_A\otimes\sigma^B),
\end{equation}
where $S_0(\rho\|\sigma)$ denotes the relative R\'enyi entropy of
order zero, of $\rho$ with respect to $\sigma$. By replacing the
relative R\'enyi entropy of order zero with the quantum relative
entropy,
\begin{equation}\label{q-rel}
S(\rho\|\sigma):=\left\{
\begin{split}
  &\Tr[\rho\log\rho-\rho\log\sigma],\textrm{ if }\supp\ \rho\subseteq\supp\ \sigma\\
  &+\infty,\textrm{ otherwise},
\end{split}
\right.
\end{equation}
we in fact obtain the usual coherent information $I^{A\to
  B}(\rho^{AB})$:
\begin{equation}\label{coh}
\begin{split}
  \min_{\sigma^B\in\states(\sH_B)}S(\rho^{AB}\|\openone_A\otimes\sigma^B)&=S(\rho^{AB}\|\openone_A\otimes\rho^B)\\
&=S(\rho^B)-S(\rho^{AB})\\
:&=I^{A\to
    B}(\rho^{AB}).
\end{split}
\end{equation}
Note in particular that for a MES of rank $M$, as defined by
\reff{MES-M}, \reff{zero-coh} yields
\begin{equation}
I^{A\to B}_{0}(\Psi_M^{AB}) = I^{A\to B}(\Psi_M^{AB})= \log
M. 
\end{equation}

Further, given an $\alpha-$relative R\'enyi entropy $S_\alpha(\rho\|\sigma)$,
for a bipartite $\rho=\rho^{AB}$, we define the corresponding
$\alpha$-conditional entropy as
\begin{equation}\label{eq:cond}
  H_\alpha(\rho^{AB}|\sigma^B):=-S_\alpha(\rho^{AB}\|\openone_A\otimes\sigma^B),
\end{equation}
and
\begin{equation}
\begin{split}
H_\alpha(\rho^{AB}|B):&=\max_{\sigma^B\in \states({\cal{H}}_B)}H_\alpha(\rho^{AB}|\sigma^B)\\
&=-\min_{\sigma^B\in\states({\cal{H}}_B)}S_\alpha(\rho^{AB}\|\openone_A\otimes\sigma^B).
\label{23}
\end{split}
\end{equation}
 Then for any $\delta >0$, the corresponding smoothed $\alpha$-conditional entropies $H^\delta_\alpha(\rho^{AB}|B)$ are 
defined as follows:
\begin{equation}\label{cond-smooth}
H_{\alpha}^\delta(\rho^{AB}|B):=\left\{
\begin{split}
  &\min_{\bar\rho^{AB}\in \B(\rho^{AB};\delta)}H_\alpha(\bar\rho^{AB}|B),\textrm{ for }0\le\alpha< 1\\
  &\max_{\bar\rho^{AB}\in \B(\rho^{AB};\delta)}H_\alpha(\bar\rho^{AB}|B),\textrm{ for }1<\alpha,\\
\end{split}
\right.
\end{equation}
and the corresponding smoothed $\alpha$-coherent information is
defined as
\begin{equation}\label{coh-smooth}
  I^{A\to B}_{\alpha,\delta}(\rho^{AB}):=-H_\alpha^\delta(\rho^{AB}|B).
\end{equation}
For $\alpha=0$, this is identical to the definition~(\ref{eq:i}).

The following lemma, proved in \cite{fbnd}, will play a central role in our proof:
\begin{lemma}[Quantum data-processing
  inequality~\cite{fbnd}]\label{lemma:data-proc}
  For any bipartite state $\rho^{AB}$, any completely positive,
  trace-preserving map $\Phi:B\mapsto C$, and any $\delta\ge 0$, we
  have
\begin{equation}\nonumber
  \I_{0,2\sqrt{\delta}}^{A\to B}(\rho^{AB})\ge\I_{0,\delta}^{A\to C}((\id\otimes \Phi)(\rho^{AB})).
\end{equation}
\end{lemma}

\section{Entanglement distillation: the ``one-shot'' case}
\label{section_main}

Let Alice and Bob, who are in two different locations, 
share a single copy of an arbitrary state $\rho^{AB}$. Their
aim is to distill entanglement from this
shared state (i.e., convert the state to a maximally entangled state) 
using local operations and classical communication (LOCC) only.
If Alice is allowed to send classical information to Bob but not allowed 
to receive any from him, then the LOCC transformation is said to be 
one-way (or forward) and is denoted by the symbol $\Lambda^\ra(\rho^{AB})$. 
More general LOCC operations in which Alice and Bob are both allowed to
send classical information to each other, are referred to as two-way LOCC and 
denoted by the symbol $\Lambda^\lra(\rho^{AB})$. 
We refer to the corresponding protocols as 
\emph{one-shot entanglement distillation} (under one-way LOCC and two-way LOCC,
respectively). Note that in a two-way LOCC, Alice
and Bob are allowed to communicate with each other classically and perform
local operations multiple times.

For sake of generality, we consider the situation where, for any given $\eps\ge 0$, the final state of the protocol is $\eps$-close to a maximally entangled state, with respect to a suitable distance measure. More precisely, we require the fidelity \reff{fidelity-aaa}
of the final state of the protocol and a maximally entangled state to be $\ge 1 - \eps$.

\begin{definition}[$\eps$-achievable distillation rates] 
  For any given $\eps\ge0$, a real number $R\ge 0$ is said to be an
  \emph{$\eps$-achievable rate} for two-way entanglement distillation,
  if there exists an integer $M\ge 2^R$ and a maximally entangled state
  $\Psi_M^{A'B'}$ such that
\begin{equation}
F\left(\Lambda^{\leftrightarrow}(\rho^{AB}),\Psi_M^{A'B'}\right)\ge 1-\eps , 
\end{equation}
for some two-way LOCC operation $\Lambda^{\leftrightarrow}:AB\mapsto
A'B'$. An analogous definition holds for one-way LOCC.
\end{definition}

\begin{definition}[One-shot distillable entanglement] For any given
  $\eps\ge0$, the one-shot distillabe entanglement, $E_D^\lra(\rho_{AB};\eps)$,  under two-way LOCC
  is the maximum of all $\eps$-achievable
  two-way entanglement distillation rates. An analogous definition
  holds for the one-shot distillable entanglement under
  one-way LOCC, which is denoted by the symbol
  $E_D^\ra(\rho_{AB};\eps)$.
 \end{definition}

\begin{definition}[One-way entanglement distillation fidelity]
  Given a bipartite state $\rho^{AB}$, for any $m\in\mathbb{N}$ we
  define the one-way entanglement distillation fidelity as follows:
\begin{equation}
  \F(\rho^{AB};m):=\max_{\Lambda^\ra}F(\Lambda^\ra(\rho^{AB}),\Psi^{A'B'}_m),
\end{equation}
where the maximization is over all one-way LOCC maps
$\Lambda^{\ra}:AB\mapsto A'B'$, and $\Psi^{A'B'}_m$ is some MES of
rank $m$. An analogous definition holds for the two-way entanglement
distillation fidelity $\mathsf{F}^{\leftrightarrow}_D(\rho^{AB};m)$.
\end{definition}

\begin{definition}[Completely positive instruments]
  A completely positive (CP) instrument~\cite{instrument} is a family
  of CP maps $\{\mE_x\}_{x \in \X}$, labelled by the parameter $x$,
  which sum up to a trace-preserving (TP) map.
\end{definition}

Roughly speaking, we can think of an instrument as a quantum operation 
with both classical and quantum outputs. We denote an instrument acting
  on a quantum system $A$ by the symbol $\iI_A:A\to A'\X$, where $A'$
  and $\X$ denote the systems corresponding to the quantum and
  classical outputs respectively. Without loss of generality, the
  action of an instrument $\iI_A$ on the state $\rho$ can be
  represented as
  $\iI_A(\rho)=\sum_{x\in\X}\mE_x(\rho)\otimes|x\>\<x|_\X$, where $|x\>$ are
  orthonormal states representing the classical register $\X$ storing
  the measurement outcome $x$.

The most general one-way entanglement distillation protocol consists 
of Alice using a CP instrument on her part of the shared bipartite state, 
communicating the classical output to Bob, and Bob performing a CPTP map 
on his part of the shared state accordingly.  

\section{One-shot bounds on distillable entanglement}\label{one_bounds}

\subsection{Lower bounds (direct parts)}

\begin{lemma}[One-Shot Hashing Bound]\label{lemma:one-hash}
  For any given bipartite state $\rho^{AB}\in \sH_A \otimes \sH_B$ 
and any $\eps\ge0$, the
  one-shot distillable entanglement via one-way (forward) LOCC
  transformations is bounded as follows:
  \begin{equation}\label{eq:hashing}
    E^\ra_D(\rho^{AB};\eps)\ge I_{0,\eps/8}^{A\to B}(\rho^{AB})+
    \log\left[\frac1{d_{A}}+\frac{\eps^2}{4}\right]-\Delta,
      \end{equation}
      where $I_{0,\eps/8}^{A\to B}$ denotes the smoothed zero-coherent
      information defined by~\reff{eq:i}, $d_A = {\rm{dim\,}} \sH_A$, and $\Delta\in[0,1]$ is a
      constant included to ensure that the right hand side
      of~(\ref{eq:hashing}) is equal to the logarithm of an integer
      number.
\end{lemma}

In order to prove the above lemma, we need the following additional
lemma, which is proved in Appendix A.
\begin{lemma}
\label{thm_one}
Given a state $\rho^{AB}$, for any $\delta\ge 0$ and any positive
integer $m\le d_A$,
  \begin{equation}\label{eq:111}
    \F(\rho^{AB};m)\ge
    1-4\delta-\sqrt{m\left\{2^{I^{A\to E}_{2,\delta}(\rho^{AE})}-\frac
        1 {d_A}\right\}},
  \end{equation}
  where $\rho^{AE}$ is the reduced state $\Tr_B[\Omega^{ABE}]$ of any arbitrary purification $|\Omega^{ABE}\>$ of $\rho^{AB}$, and $I^{A\to
    E}_{2,\delta}(\rho^{AE})$ is given by~\reff{coh-smooth} for
  $\alpha=2$.
\end{lemma}

\begin{proof}[Proof of Lemma~\ref{lemma:one-hash}]
For any fixed $\varepsilon \ge 0$, a positive real number $R=\log m$ is an $\eps$-achievable rate for one-way
  distillation if $\F(\rho^{AB},m)\ge 1-\eps$. Due to
  Eq.~(\ref{eq:111}), we know that $R=\log m$ is achievable if
  \begin{equation}
    4\delta+\sqrt{m\left\{2^{I^{A\to E}_{2,\delta}(\rho^{AE})}-\frac
        1 {d_A}\right\}}\le\eps.
  \end{equation}
  For $0\le\delta\le\eps/4$, $\log m$ is achievable if, in particular,
\begin{equation}\label{eq:asdf}
  m2^{I^{A\to E}_{2,\delta}(\rho^{AE})}\le(\eps-4\delta)^2+\frac 1{d_A},
\end{equation}
since $m/d_A\ge 1/d_A$. Since $E_D^\ra(\rho^{AB};\eps)$ is defined as
the maximum over all achievable rates, Eq.~(\ref{eq:asdf}) implies
that, for all $\delta\in[0,\eps/4]$,
\begin{equation}\label{27}
  E_D^\ra(\rho^{AB};\eps)\ge\log\left[(\eps-4\delta)^2+\frac 1{d_A}\right]-I^{A\to E}_{2,\delta}(\rho^{AE})-\Delta,
\end{equation}
where $\Delta$ is a positive number, less than or equal to one,
subtracted in order to make the right hand side of the above equation
equal the logarithm of an integer number (as it has to be, by
definition).

The last ingredient needed to complete the proof of Lemma~\ref{lemma:one-hash}
is the fact that, for the reduced states $\rho^{AB}$ and
$\rho^{AE}$ of the same pure state $|\Omega^{ABE}\>$, 
\be\label{28}
-I^{A\to
  E}_{2,\delta}(\rho^{AE})\ge I^{A\to B}_{0,\delta}(\rho^{AB}).
\ee
This inequality is stated as Lemma~\ref{i2i0} of Appendix B,
where it is proved using duality arguments along the lines
following~\cite{roger}. The statement of
Lemma~\ref{lemma:one-hash} is finally obtained from \reff{27}
and \reff{28} for $\delta=\eps/8$.
\end{proof}

Due to the one-shot hashing bound, Lemma~\ref{lemma:one-hash}, we know
that the zero-coherent information is an achievable rate for one-way
entanglement distillation. Since the zero-coherent information can in
general increase under the action of an LOCC transformation, we can
think of pre-processing the initial state by means of a suitable LOCC
map, and distilling entanglement out of the pre-processed state,
instead of the initial given one. This procedure leads us to the
following achievable rates for one- and two-way entanglement
distillation:

\begin{corollary}[Lower bounds]\label{lower_bound}
  Let $\iI_A:A\to A'\X$ denote an instrument on $A$, and let
  $\Lambda^{\ra}_{AB}:AB\to A'B'$ denote a one-way (from $A$ to $B$)
  LOCC transformation. Then,
  \begin{equation}
\begin{split}
  E_D^\ra(\rho^{AB};\eps)&\ge
  \max_{\Lambda_{AB}^{\ra}}I_{0,\eps/8}^{A'\to
    B'}(\sigma^{A'B'})+\log\left[\frac
    1{d_{A'}}+\frac{\eps^2}{4}\right]-\Delta\\
  &\ge\max_{\iI_A}I_{0,\eps/8}^{A'\to
    B\X}(\sigma^{A'B\X})+\log\left[\frac
    1{d_{A'}}+\frac{\eps^2}{4}\right]- \Delta' \label{lb},
\end{split}
\end{equation}
where $\sigma^{A'B'}=\Lambda_{AB}^{\ra}(\rho^{AB})$,
$\sigma^{A'B\X}=(\iI_A\otimes\id_B)(\rho^{AB})$, and
$\Delta,\Delta'\in[0,1]$ are included to ensure that the lower bounds
\reff{lb} are each equal to the logarithm of a positive integer.

Analogously, let $\Lambda^{\lra}_{AB}:AB\to A'B'$ be a two-way LOCC
transformation. Then,
\begin{equation}
E_D^\lra(\rho^{AB};\eps)\ge \max_{\Lambda_{AB}^{\lra}}I_{0,\eps/8}^{A'\to
    B'}(\sigma^{A'B'})+\log\left[\frac
    1{d_{A'}}+\frac{\eps^2}{4}\right]-\Delta'',
\end{equation}
where $\Delta''\in[0,1]$ is included to ensure that the lower bound is
equal to the logarithm of a positive integer.
\end{corollary}

\subsection{Upper bounds (converse parts)}

When distilling entanglement with one-way LOCC protocols, there is no
need to employ a full one-way LOCC transformation when pre-processing
the initial state: the following lemma shows that in fact an
instrument on Alice's side only, followed by the communication of the
outcome to Bob, suffices.

\begin{lemma}[One-way weak converse]\label{lemma:one-conv}
For any given bipartite state $\rho^{AB}$ and any $\eps\ge0$,
\begin{equation}\label{eq:one-way}
  E_D^\ra(\rho^{AB};\eps)\le\max_{\iI_A}\I_{0,4\sqrt{\eps}}^{A'\to
    B\X}(\sigma^{A'B\X}),
\end{equation}
where the maximization is done over instruments
$\iI_A:=\{\mE^m\}_{m\in\X}$, where each $\mE^m$ maps $A$ to $A'$, and
$\sigma^{A'B\X}:=
\sum_m(\mE^m\otimes\id_B)(\rho^{AB})\otimes|m\>\<m|^\X$.
\end{lemma}

\begin{proof}
  Suppose that $R$ is a one-way $\eps$-achievable rate, i.~e. there
  exists an integer $M$ with $\log M\ge R$ and a one-way forward LOCC
  operation $\Lambda^\ra:AB\mapsto A'B'$ such that
\begin{equation}
  \<\Psi^{A'B'}_M|\Lambda^\ra(\rho^{AB})|\Psi^{A'B'}_M\>\ge(1-\eps)^2.
\end{equation}
The most general one-way forward LOCC operation is constructed as
follows: $(i)$ Alice applies a CP instrument on her share, $(ii)$ she
communicates the outcome $m$ to Bob, $(iii)$ Bob deterministically
performs a decoding operation on his share, depending on Alice's
outcome. Such a procedure is conveniently represented by writing the
following classical-quantum (c-q) state
\begin{equation}
  \tau^{A'B'\X}:=
\sum_{m\in\X}(\mE^m\otimes\mD^m)(\rho^{AB})\otimes|m\>\<m|^\X,
\end{equation}
where the $\mD^m$'s are CPTP maps for all $m$, while the $\mE^m$'s
are just CP maps normalized so that their sum $\mE:=\sum_m\mE^m$ is
TP. The classical flags $|m\>\<m|^\X$ represent the classical
information that Alice sends to Bob. This is the reason why we
consider both systems $B'$ and $\X$ to be in Bob's hands.

By the quantum data-processing inequality, Lemma~\ref{lemma:data-proc}, we
have that
\begin{equation}\label{eq:334}
\I_{0,\delta}^{A'\to B'\X}(\tau^{A'B'\X})\le \widetilde{I}_{0,2\sqrt{\delta}}^{A'\to
  B\X}(\sigma^{A'B\X}),
\end{equation}
where $\sigma^{A'B\X}:=
\sum_m(\mE^m\otimes\id_B)(\rho^{AB})\otimes|m\>\<m|^\X$. Moreover,
since we assumed that $\<\Psi_M^{A'B'}|\tau^{A'B'}|\Psi_M^{A'B'}\>\ge
(1-\eps)^2$, the operator
$$P:=\sum_m|\Psi_M\>\<\Psi_M|^{A'B'}\otimes|m\>\<m|^\X$$ is such that
$\Tr[P\ \tau^{A'B'\X}]\ge (1-\eps)^2\ge 1-2\eps$. Then, continuing
from~(\ref{eq:334}), and recalling the definition in~(\ref{eq:itilda}),
\begin{equation}
\begin{split}
\widetilde{I}_{0,4\sqrt{\eps}}^{A'\to
  B\X}(\sigma^{A'B\X})&\ge
\I_{0,2\eps}^{A'\to
  B'\X}(\tau^{A'B'\X})\\
&\ge-\max_{\omega^{B'\X}}\log\Tr[\sqrt{P}\Pi_{\tau^{A'B'\X}}\sqrt{P}\
(\openone_{A'}\otimes\omega^{B'\X})]\\
&\ge-\max_{\omega^{B'\X}}\log\Tr[P\
(\openone_{A'}\otimes\omega^{B'\X})]\\
&=I_0^{A'\to B'\X}\left(\sum_mq_m|\Psi_M\>\<\Psi_M|^{A'B'}\otimes|m\>\<m|^\X\right),
\end{split}
\end{equation}
for any probability distribution $q_m>0$, $\sum_mq_m=1$. Finally,
since the quantum data-processing inequality also holds when the 
smoothing parameter is equal to zero, we have
\begin{equation}
\begin{split}
I_0^{A'\to
  B'\X}\left(\sum_mq_m|\Psi_M\>\<\Psi_M|^{A'B'}\otimes|m\>\<m|^\X\right)&\ge I_0^{A'\to B'}(\Psi_M^{A'B'})\\
&=\log M\\
&\ge R.
\end{split}
\end{equation}
Hence we have proved that, if $R$ is an $\eps$-achievable rate, there
always exists an instrument $\iI=\{\mE^m\}_{m\in\X}$ on $A$ such that
\begin{equation}
\widetilde{I}_{0,4\sqrt{\eps}}^{A'\to
  B\X}(\sigma^{A'B\X})\ge R.
\end{equation}
This in turn implies that 
$$R \le\max_{\iI_A}\I_{0,4\sqrt{\eps}}^{A'\to
    B\X}(\sigma^{A'B\X}).$$
Then (\ref{eq:one-way}) is obtained by taking the maximum over all
$\eps$-achievable rates.
\end{proof}

While for the one-way distillation scenario the pre-processing can be
reduced, without loss of generality, to an instrument at Alice's side
only, in the two-way scenario we have to keep the pre-processing as
general as possible. In particular, for any $\eps \ge 0$, we obtain
the following upper bound to the one-shot distillable entanglement
under two-way LOCC transformations.
\begin{lemma}\label{lemma:two-conv}
For any given bipartite state $\rho^{AB}$ and any $\eps\ge0$, the 
distillable entanglement under two-way LOCC satisfies the following bound:
\begin{equation}
  E_D^{\lra}(\rho^{AB};\eps)\le\max_{\Lambda^{\lra}_{AB}}\I_{0,{2\eps}}^{A'\to
    B'}(\omega^{A'B'}),
\end{equation}
where the maximization is done over two-way LOCC transformations
$\Lambda^\lra_{AB}$ mapping $AB$ to $A'B'$, and $\omega^{A'B'}:=
\Lambda_{AB}^\lra(\rho^{AB})$.
\end{lemma}
\begin{proof}
  Let $\Lambda_{AB}^\lra$ be a two-way LOCC transformation whose
  action on the state $\rho^{AB}$ yields the state $\omega^{A'B'}:=
  \Lambda_{AB}^\lra(\rho^{AB})$, such that $F(\omega^{A'B'},
  \Psi_M^{A'B'}) \ge 1-\eps$, and $R = \log M$.

By the definitions~\reff{eq:itilda} and~\reff{zero-coh} of the zero
coherent information, we have
\begin{equation}
\begin{split}
  R=\log M &= \min_{\sigma^B\in\states(\sH_B)}\Bigl[- \log \tr\Bigl(\Psi_M^{A'B'}(\openone_{A'}\otimes\sigma^{B'})\Bigr) \Bigr]\\
  &\le \min_{\sigma^{B'}\in\states(\sH_{B'})}\Bigl[- \log
  \tr\Bigl(\Psi_M^{A'B'} \Pi_{\omega^{A'B'}}\Psi_M^{A'B'}
  (\openone_{A'}\otimes\sigma^{B'})\Bigr) \Bigr]\\
  &\le \max_{P\in \P(\omega^{A'B'};2\eps)}\min_{\sigma^{B'}\in\states(\sH_{B'})}
  \Bigl[- \log \tr\Bigl(\sqrt{P} \Pi_{\omega^{A'B'}}\sqrt{P}
  (\openone_{A'}\otimes\sigma^{B'})\Bigr) \Bigr]\\
  &=\I_{0,2\eps}^{A'\to B'}(\omega^{A'B'})\\
  &\le \max_{\Lambda^{\lra}_{A'B'}}\I_{0,{2\eps}}^{A'\to
    B'}(\omega^{A'B'}),
\end{split}
\end{equation}
where the second identity follows from the definition~\reff{zero-coh}
of the zero coherent information, and the fact that $\Pi_{\Psi_M^{A'B'}}
= \Psi_M^{A'B'}$; the first inequality follows from
$\Pi_{\omega^{A'B'}}\le \openone$, and the second inequality follows
from the fact that $\Psi_M^{A'B'} \in \P(\omega^{A'B'};2\eps)$ (see
definition~(\ref{P-ball})) since $\tr[\Psi_M^{A'B'}\omega^{A'B'}]\ge 1 -
2\eps$, which in turn follows from the fact that $F^2(\omega^{A'B'},
\Psi_M^{A'B'}) \ge (1-\eps)^2$.
\end{proof}

\section{Main result: exact asymptotic formulas for arbitrary
  resources}\label{asymp}
In this section we consider entanglement distillation from arbitrary
resources, comprising an arbitrary sequence of bipartite states
$\hat\rho_{AB}:=\{\rho_{AB}^n\}_{n=1}^\infty$, where $\rho_{AB}^n \in
\states(\sH_A^{\otimes n}\otimes \sH_B^{\otimes n})$.  The one-way
distillable entanglement rate for such a sequence is defined as:
\begin{equation}
E_{D,\infty}^\ra(\hat\rho_{AB}):=\lim_{\eps\to
  0}\liminf_{n\to\infty}\frac 1n E_D^\ra(\rho_{AB}^n;\eps),
\end{equation}
and the two-way distillable entanglement
$E_{D,\infty}^\lra(\hat\rho_{AB})$ is defined analogously.

To evaluate the distillable entanglement of such a sequence of
states, we employ the well-known Quantum Information Spectrum
Method~\cite{info-spec,hayashi-naga}. Two fundamental quantities used
in this approach are the \emph{quantum spectral sup}- and
\emph{inf-divergence rates}, defined as follows:
\begin{definition}[Spectral Divergence Rates]
  Given a sequence of states $\hat\rho=\{\rho_n\}_{n=1}^\infty$ and a
  sequence of positive operators
  $\hat\sigma=\{\sigma_n\}_{n=1}^\infty$, the quantum spectral
  sup- (inf-)divergence rates are defined in terms of the difference
  operators $\Pi_n(\gamma) = \rho_n - 2^{n\gamma}\sigma_n$ as
\begin{align}
  \overline{D}(\hat\rho \| \hat\sigma) &:= \inf \left\{ \gamma : \limsup_{n\rightarrow \infty} \mathrm{Tr}\left[ \{ \Pi_n(\gamma) \geq 0 \} \Pi_n(\gamma) \right] = 0 \right\} \label{odiv} \\
  \underline{D}(\hat\rho \| \hat\sigma) &:= \sup \left\{ \gamma :
  \liminf_{n\rightarrow \infty} \mathrm{Tr}\left[ \{ \Pi_n(\gamma) \geq
  0 \} \Pi_n(\gamma) \right] = 1 \right\} \label{udiv}
\end{align}
respectively.  
\end{definition}

It is known that (see e.g.~\cite{bowen-datta})
\begin{equation}\label{rell}
  \overline{D}(\hat\rho \| \hat\sigma) \ge\lim_{n\to\infty}\frac 1nS(\rho_n\|\sigma_n)\ge\underline{D}(\hat\rho \| \hat\sigma). 
\end{equation}

In analogy with the usual definition of the coherent
information~\reff{coh}, we moreover define the \emph{spectral sup-}
and \emph{inf-coherent information rates}, respectively, as follows:
\begin{align}
  \overline{I}^{A\to B}(\hat\rho_{AB})&:=\min_{\hat\sigma_B}\overline{D}(\hat\rho_{AB}\|\hat\openone_A\otimes\hat\sigma_B),\label{strong-conv}\\
  \underline{I}^{A\to B}(\hat\rho_{AB})&:=\min_{\hat\sigma_B}\underline{D}(\hat\rho_{AB}\|\hat\openone_A\otimes\hat\sigma_B),\label{att}
\end{align}
where $\hat\rho_{AB}:=\{\rho^n_{AB}\in\states(\sH_A^{\otimes
  n}\otimes\sH_B^{\otimes n})\}_{n=1}^\infty$,
$\hat\sigma_B:=\{\sigma_B^n\in\states(\sH_B^{\otimes
  n})\}_{n=1}^\infty$, and $\hat\openone_A:=\{ \openone_A^{\otimes
  n}\}_{n=1}^\infty$. The inequality~\reff{rell} ensures that
\begin{equation}
  \overline{I}^{A\to B}(\hat\rho_{AB}) \ge\lim_{n\to\infty}\frac 1n 
I^{A\to B}(\rho_{AB}^n)\ge\underline{I}^{A\to B}(\hat\rho_{AB}).
\end{equation}
Note that in eq.~\reff{strong-conv} and~\reff{att} we could write
minimum instead of infimum due to Lemma~1 of~\cite{hayashi-naga}.

Let $\hiI_A:= \{\iI_A^n\}_{n=1}^\infty$ denote a sequence of
instruments $\iI_A^n:A_n\to A^\prime_n\X_n$. Our main results on the
distillable entanglement for an arbitrary sequence of states
$\hat\rho_{AB}$ under both one-way LOCC and two-way LOCC, are given by
the following theorem.\medskip

\framebox[\linewidth]{
\begin{minipage}{0.95\linewidth}
\begin{theorem}\label{thm:main}
  Given a sequence of bipartite states $\hat\rho_{A
    B}:=\{\rho_{AB}^n\}_{n=1}^\infty$,
  \begin{equation}
    E_{D,\infty}^\ra(\hat\rho_{ A B})=\max_{\hiI_{
        A}}\underline{I}^{A'\to B\X}(\hat\sigma_{ A'\X
    B}),\label{17}
  \end{equation}
and
\begin{equation}
    E_{D,\infty}^\lra(\hat\rho_{ AB})=\max_{\hat\Lambda^{\lra}_{
        A B}}\underline{I}^{ A'\to  B'}(\hat\nu_{ A'
    B'}),\label{18}
  \end{equation}
  where: the maximisation in~\reff{17} is over all sequences of
  instruments, $\hiI_{A}:= \{\iI_A^n\}_{n=1}^\infty$, and
  $\hat\sigma_{ A'\X B}:= \{\iI_A^n(\rho_{AB}^n)\}_{n=1}^\infty$; the
  maximisation in~\reff{18} is over all sequences of two-way LOCC
  operations, $\hat\Lambda^{\lra}_{A B}:= \{\Lambda_{A
    B}^n\}_{n=1}^\infty$, and $\hat\nu_{ A'B'}:= \{{\Lambda^n_{A
      B}}(\rho_{AB}^n)\}_{n=1}^\infty$.
\end{theorem}
\end{minipage}
}
\bigskip

From Corollary~\ref{lower_bound} and Lemma~\ref{lemma:one-conv}, we
have that, for any $\eps >0$ and any $n\ge 1$,
\begin{equation}
\begin{split}
  &\frac 1n\max_{\iI_A^n}\I_{0,4\sqrt{\eps}}^{A'\to
    B\X}(\sigma_{A'B\X}^n)\\
  \ge
  &\frac{1}{n} E_D^\ra(\rho_{AB}^n;\eps)\\
  \ge &\frac{1}{n} \max_{\iI_A^n}I_{0,\eps/8}^{A'\to
    B\X}(\sigma_{AB\X}^n)+\frac{1}{n}\log\left[\frac
    1{d_{A'_n}}+\frac{\eps^2}{4}\right]-
  \frac{\Delta'}{n}, \label{lowb1}
\end{split}
\end{equation}
where $\sigma^n_{A'B\X}=(\iI^n_A\otimes\id_B)(\rho^n_{AB})$.  In the
case of two-way entanglement distillation, again
Corollary~\ref{lower_bound} and Lemma~\ref{lemma:two-conv} yield
\begin{equation}
\begin{split}
  &\frac 1n\max_{\Lambda_{AB}^n}\I_{0,2\eps}^{A'\to
    B'}(\nu_{A'B'}^n)\\
  \ge
  &\frac{1}{n} E_D^\lra(\rho_{AB}^n;\eps)\\
  \ge &\frac{1}{n} \max_{\Lambda_{AB}^n}I_{0,\eps/8}^{A'\to
    B'}(\nu_{A'B'}^n)+\frac{1}{n}\log\left[\frac
    1{d_{A'_n}}+\frac{\eps^2}{4}\right]-
  \frac{\Delta''}{n}, \label{lowb2}
\end{split}
\end{equation}

Theorem~\ref{thm:main} then follows rather straightforwardly by taking
the limits $\lim_{\eps\to 0}\lim_{n\to\infty}$ on either sides 
of the inequalities~\reff{lowb1} and~\reff{lowb2}, and applying the following
two lemmas (which were proved in~\cite{fbnd}):

\begin{lemma}[Direct part~\cite{fbnd}]\label{lemma:x}
  Given a sequence of bipartite states $\hat\rho_{AB}$,
\begin{equation}\nonumber
\begin{split}
  \lim_{\delta\to
    0}\liminf_{n\to\infty}&\max_{\bar\rho_{AB}^n\in\B(\rho_{AB}^n;\delta)}\min_{\sigma_{B}^n}\frac
  1nS_0(\bar\rho^n_{AB}\|\openone_{A}^{\otimes n}\otimes\sigma_{B}^n)\\
  &\ge\min_{\hat\sigma_B}\underline{D}(\hat\rho_{AB}\|\hat\openone_A\otimes\hat\sigma_B),
\end{split}
\end{equation}
or, equivalently, $\lim_{\delta\to 0}\liminf_{n\to\infty}\frac 1n I_{0,\delta}^{A\to B}(\rho^n_{AB})\ge\underline{I}^{A\to B}(\hat\rho_{AB})$.
\end{lemma}

\begin{lemma}[Weak converse~\cite{fbnd}]\label{S0Dub}
  Given a sequence of bipartite states $\hat\rho_{AB}$,
\begin{equation}\nonumber
\begin{split}
  \lim_{\delta\to
    0}\liminf_{n\to\infty}&\max_{P_n\in\P(\rho_{AB}^n;\delta)}\min_{\sigma_{B}^n}\frac
  1nS_0^{P_n}(\rho^n_{AB}\|\openone_{A}^{\otimes n}\otimes\sigma_{B}^n)\\
&\le\min_{\hat\sigma_B}\underline{D}(\hat\rho_{AB}\|\hat\openone_A\otimes\hat\sigma_B),
\end{split}
\end{equation}
or, equivalently, $\lim_{\delta\to 0}\liminf_{n\to\infty}\frac 1n \I_{0,\delta}^{A\to B}(\rho^n_{AB})\le\underline{I}^{A\to B}(\hat\rho_{AB})$.
\end{lemma}

\subsection{The special case of i.i.d. resources}

Let us now consider the case in which Alice and Bob share
multiple, independent and identical copies of a given bipartite
state $\rho_{AB} \in \states(\sH_A \otimes \sH_B)$. 
The entanglement resource is in this case characterized by 
the sequence $\hrho_{AB}:=\{\rho_{AB}^{\otimes n}\}_{n=1}^\infty$.
The asymptotic distillable entanglement of the state $\rho_{AB}$
can be obtained from Theorem~\ref{thm:main} by employing the 
following lemma, which was proved in~\cite{cost} by using the 
Generalized Stein's Lemma~\cite{brandao-plenio}.
\begin{lemma}\label{lemma:seven}
  For any given bipartite state $\rho_{AB}$
\begin{equation}
  \min_{\hat\sigma_B}\underline{D}(\hat\rho_{AB}\|\hat\openone_A\otimes\hat\sigma_B)=S(\rho_{AB}\|\openone_A\otimes
  \rho_B),
\end{equation}
where $\hat\rho_{AB}=\{\rho_{AB}^{\otimes n}\}_{n=1}^\infty$,
$\hat\sigma_B:=\{\sigma_B^n\in\states(\sH_B^{\otimes
  n})\}_{n=1}^\infty$, and $\hat\openone_A:=\{\openone_A^{\otimes
  n}\}_{n=1}^\infty$. Notice that the optimizing sequence
$\hat\sigma_B$ is not i.i.d. in general.
\end{lemma}

We can then retrieve the expressions for the asymptotic 
distillable entanglement of any arbitrary bipartite state $\rho^{AB}$,
obtained in~\cite{devetak-winter}, as a corollary
of our Theorem~\ref{thm:main}:\medskip

\framebox[\linewidth]{
\begin{minipage}{0.95\linewidth}
\begin{corollary}[\cite{devetak-winter}]
  For any bipartite state $\rho^{AB}$, the one-way distillable
  entanglement rate is given by
\begin{equation}
  E_{D,\infty}^\ra(\rho^{AB})=\lim_{n\to\infty}\frac
  1n\max_{\iI^n_A}I^{A'\to B\X}(\sigma_{A'B\X}^n),
\end{equation}
where $\sigma_{A'B\X}^n=(\iI^n_A\otimes\id_B)(\rho_{AB}^{\otimes
  n})$. The two-way distillable entanglement rate is given by
\begin{equation}
  E_{D,\infty}^\lra(\rho^{AB})=\lim_{n\to\infty}\frac
  1n\max_{\Lambda^n_{AB}}I^{A'\to B'}(\nu_{A'B'}^n),
\end{equation}
where $\nu_{A'B'}^n=\Lambda^n_{AB}(\rho_{AB}^{\otimes n})$.
\end{corollary}
\end{minipage}
}

\begin{proof}
  Let $\hat\rho_{AB}$ be the i.i.d. sequence $\{\rho_{AB}^{\otimes
    n}\}_{n=1}^\infty$. From Theorem~\ref{thm:main}, we have that
  \begin{equation}
    E_{D,\infty}^\ra(\hat\rho_{AB})\ge\max_{\iI_A}\underline{I}^{A'\to B\X}(\hat\sigma_{A'B\X}),
  \end{equation}
  where $\sigma^n_{A'B\X}=(\iI_A^{\otimes n}\otimes\id_B^{\otimes
    n})(\rho_{AB}^{\otimes n})$, i.e., the sequence
  $\hat\sigma_{A'B\X}$ is i.i.d. Due to Lemma~\ref{lemma:seven} then,
  \begin{equation}
    E_{D,\infty}^\ra(\hat\rho_{AB})\ge\max_{\iI_A}I^{A'\to B\X}(\sigma_{A'B\X}).
  \end{equation}
By a standard blocking argument, it follows that, in particular, for any $m\ge 1$,
\begin{equation}
   E_{D,\infty}^\ra(\rho^{AB})\ge\frac
  1m\max_{\iI^m_A}I^{A'\to B\X}(\sigma_{A'B\X}^m),
\end{equation}
where $\sigma^m_{A'B\X}=(\iI_A^m\otimes\id_B^{\otimes
  m})(\rho_{AB}^{\otimes m})$. By taking the limit $m\to\infty$, we
obtain
\begin{equation}
  E_{D,\infty}^\ra(\rho^{AB})\ge\lim_{m\to\infty}\frac
  1m\max_{\iI^m_A}I^{A'\to B\X}(\sigma_{A'B\X}^m).
\end{equation}
The converse direction, that is,
\begin{equation}
  E_{D,\infty}^\ra(\rho^{AB})\le\lim_{m\to\infty}\frac
  1m\max_{\iI^m_A}I^{A'\to B\X}(\sigma_{A'B\X}^m),
\end{equation}
simply comes from the fact that the
$\underline{D}(\hat\rho\|\hat\sigma)\le\lim_{n\to\infty}\frac
1nS(\rho^n\|\sigma^n)$.

The proof for $E_{D,\infty}^\lra(\rho^{AB})$ follows from exactly the
same line of arguments.
\end{proof}

\section*{Acknowledgments}

FB acknowledges support from the Program for Improvement of Research
Environment for Young Researchers from Special Coordination Funds
for Promoting Science and Technology (SCF) commissioned by the Ministry
of Education, Culture, Sports, Science and Technology (MEXT) of Japan.
ND acknowledges support from the European Community's Seventh Framework Programme
(FP7/2007-2013) under grant agreement number 213681. This work was completed when FB was visiting the Statistical Laboratory of the University of Cambridge.

\appendix

\section*{Appendix A: Proof of Lemma~\ref{thm_one}}

The following lemmas are employed in the proof of Lemma ~\ref{thm_one}. 
\begin{lemma}[\cite{thesis}, \cite{fbnd}]\label{lemma:tr-hs}
  For any self-adjoint operator $X$ and any positive operator $\xi>0$,
  we have
  \begin{equation}\label{orig}
    \N{X}_1^2 \le\Tr[\xi]\Tr\left[X\xi^{-1/2}X\xi^{-1/2}\right]
\le\Tr[\xi]\Tr\left[X^2\xi^{-1}\right].\ 
\end{equation}
\end{lemma}

\begin{lemma}\label{lemma:krs}
  Given a tripartite pure state $|\Omega^{A'BE}\>\in\sH_{A'}\otimes
  \sH_B\otimes \sH_E$, let $\omega^{A'B}$, $\omega^{A'E}$, and
  $\omega^E$ be its reduced states. Then, for any state $\chi^E$,
\begin{equation}
  \max_{\mathcal{D}}F^2((\id_{A'}\otimes\mD_B)(\omega^{A'B}),\Psi_m^{A'B'})
  \ge F^2(\omega^{A'E},\tau^{A'}\otimes\chi^E),
\end{equation}
where $|\Psi_m^{A'B'}\>\in\sH_{A'}\otimes\sH_{B'}$ is some fixed
maximally entangled state of rank $m$,
$\tau^{A'}=\Tr_{B'}[\Psi_m^{A'B'}]$, and
$\mathcal{D}:\mathcal{B}(\sH_B)\mapsto\mathcal{B}(\sH_{B'})$ denotes a
completely positive, trace-preserving (CPTP) map. The same holds also if the norm of the vector
$|\Omega^{A'BE}\>$ is not normalized to one.
\end{lemma}

\begin{proof} Fix some purification $|\chi^{RE}\>\in\sH_R\otimes\sH_E$
  of $\chi^E$. Then, for the fixed purification $|\Psi_m^{A'B'}\>$ of
  $\tau^{A'}$, we have, by Uhlmann's theorem~\cite{uhlmann}, the
  monotonicity of the fidelity under partial trace, and Stinespring's
  Dilation Theorem~\cite{stine},
\begin{equation}\label{eq:proof_4}
  \begin{split}
    &F^2(\omega^{A'E},\tau^{A'}\otimes\chi^E)\\
    =&\max_{|\varphi^{A'B'RE}\>\atop{\Tr_{B'R}[\varphi^{A'B'RE}]=\omega^{A'E}}}F^2(\varphi^{A'B'RE},\Psi_m^{A'B'}\otimes\chi^{RE})\\
    =&\max_{V:B\to B'R\atop{V^\dag V=\openone_B}}F^2\left((\openone^{A'}\otimes V_B\otimes \openone^E)\Omega^{A'BE}(\openone^{A'}\otimes V_B^\dag\otimes\openone^E),\Psi_m^{A'B'}\otimes\chi^{RE}\right)\\
    \le&\max_{\mD}F^2\left((\id_{A'}\otimes\mD_B)(\omega^{A'B}),\Psi_m^{A'B'}\right),
\end{split}
\end{equation}
where $\mD:\mathcal{B}(\sH_B)\mapsto\mathcal{B}(\sH_{B'})$ denotes a
CPTP map. In the second equality of~(\ref{eq:proof_4}) we used
the fact that all possible purifications of a given mixed
state ($\omega^{A'E}$, in our case) are related by some local isometry
acting on the purifying system only (i.e. subsystem $B$).\end{proof}

\begin{lemma}\label{lemma:genfid}
  For any $P,Q\ge 0$,
\begin{equation}
  F(P,Q):=\N{\sqrt{P}\sqrt{Q}}_1\ge\frac{\Tr P+\Tr Q}2-\frac 12\N{P-Q}_1.
\end{equation}
\end{lemma}

\begin{proof}
By adapting the proof in, e.g., Ref.~\cite{nielsen}, we see that
\begin{equation}
  F(P,Q)=\min_{\{E_m\}:\textrm{POVM}}\sum_m\sqrt{p_m}\sqrt{q_m},
\end{equation}
where $p_m:=\Tr[E_mP]$, and $q_m:=\Tr[E_mQ]$. Also,
\begin{equation}
  \N{P-Q}_1=\max_{\{E_m\}:\textrm{POVM}}\sum_m|p_m-q_m|.
\end{equation}
Again according with Ref.~\cite{nielsen}, let $\{\bar E_m\}$ be the
POVM achieving $F(P,Q)$, and let $\bar p_m$ and $\bar q_m$ be the
corresponding coefficients. Then,
\begin{equation}
  \begin{split}
\N{P-Q}_1&\ge\sum_m|\bar p_m-\bar q_m|\\
&=\sum_m|\sqrt{\bar p_m}-\sqrt{\bar q_m}|\cdot|\sqrt{\bar
  p_m}+\sqrt{\bar q_m}|\\
&\ge\sum_m\left(\sqrt{\bar p_m}-\sqrt{\bar q_m}\right)^2\\
&=\sum_m\bar p_m+\sum_m\bar q_m-2\sum_m \sqrt{\bar p_m}\sqrt{\bar
  q_m}\\
&=\Tr P+\Tr Q-2F(P,Q).
\end{split}
\end{equation}
\end{proof}

\begin{proof}[Proof of Lemma~\ref{thm_one}]
  The most general transformation composed of local operations and
  forward classical communication (one-way LOCC) can be written as
\begin{equation}
 \Lambda^\ra(\rho^{AB})=\int(\mE_\mu\otimes\mD_\mu)(\rho^{AB})\d\mu,
\end{equation}
where $\d\mu$ is an appropriate measure, the $\mD_\mu:B\to B'$ are
CPTP maps for all $\mu$, while the $\mE_\mu:A\to A'$ are completely
positive (CP) maps normalized so that $\mE:=\int\mE_\mu\d\mu$ is
trace-preserving (TP). The physical interpretation of such a
transformation is that, $(i)$ Alice performs a measurement on her
share, $(ii)$ she communicates the outcome $\mu$ to Bob, $(iii)$ Bob
deterministically performs a decoding operation on his share,
depending on Alice's outcome.

In the following, we will construct one particular one-way LOCC and
evaluate how good that is for distilling entanglement. Let us fix the
value of the positive integer $m\le d_A$ and define
  \begin{equation}
    \mE_g(\rho^A):=\frac {d_A}mP_m^AU_g^A\rho^A(U_g^A)^\dag (P_m^A)^\dag,
  \end{equation}
  where $U_g^A$ is a unitary representation of the element $g$ of the
  group $\mathbb{SU}(d_A)$ and
\begin{equation}\label{Pa}
P^A_m=\sum_{i=1}^m|i^{A'}\>\<i^A|,
\end{equation}
the vectors $|i^A\>$, $i=1,\ldots,d_A$, being the same as in
eq.~(\ref{MES-M}). Then, by introducing the Haar measure $\d g$ on
$\mathbb{SU}(d_A)$, it is a standard calculation to check that
\begin{equation}
\begin{split}
  \int\mE_g(\rho^A)\d g&=\frac {d_A}mP_m^A\left(\int
    U_g^A\rho^A(U_g^A)^\dag\d g\right) (P_m^A)^\dag\\
  &=\frac {d_A}mP_m^A\frac{\openone^A}{d_A}(P_m^A)^\dag\\
&=\frac{P_m^A(P_m^A)^\dag}{m},
\end{split}
\end{equation}
for all states $\rho^A$, i.e., the average map is trace-preserving.

For later convenience, starting from a fixed pure state
$|\Omega^{ABE}\>$ purifying $\rho^{AB}$, let us define the
unnormalized state
\begin{equation}\nonumber
  |\Omega_{m,g}^{A'BE}\>:=\sqrt{\frac {d_A}m}(P_m^AU_g^A\otimes\openone_B\otimes\openone_E)|\Omega^{ABE}\>.
\end{equation}
The reduced unnormalized states $\Tr_E[\Omega^{A'BE}_{m,g}]$ and
$\Tr_B[\Omega^{A'BE}_{m,g}]$ will be correspondingly denoted as
$\omega^{A'B}_{m,g}$ and $\omega^{A'E}_{m,g}$ (and so on).

By definition, the one-way distillation fidelity $\F(\rho^{AB},m)$
satisfies the bound:
\begin{equation}\label{avg}
\begin{split}
  \F(\rho^{AB},m)\ge&F\left(\int
  \max_{\mD}(\id_{A'}\otimes\mD_B)(\omega^{A'B}_{m,g})\d g,\Psi^{A'B'}_m\right)\\
\ge&\int\d g\ p(m,g)
  \max_{\mD}F\left((\id_{A'}\otimes\mD_B)(\tilde\omega^{A'B}_{m,g}),\Psi^{A'B'}_m\right),
\end{split}
\end{equation}
where the second line comes from concavity of the fidelity, and
$p(m,g):=\Tr \omega^{A'B}_{m,g}$. In Eq.~(\ref{avg}),
$|\Psi^{A'B'}_m\>$ is any MES of rank $m$ purifying
$\tau^{A'}_m:=\frac 1mP_m^A(P_m^A)^\dag$.

Using Lemma~\ref{lemma:krs}, we have
\begin{equation}\nonumber
\begin{split}
\F(\rho^{AB},m) &\ge \int\d g\ p(m,g)
F\left(\tilde\omega^{A'E}_{m,g},\tau^{A'}_m\otimes\tilde\omega^E_{m,g}\right)\\
&=\int\d g\ 
F\left(\omega^{A'E}_{m,g},\tau^{A'}_m\otimes\omega^E_{m,g}\right),
\end{split}
\end{equation}
where, in the second line, we used the fact that
$F(p\rho,p\sigma)=pF(\rho,\sigma)$. Further, using
Lemma~\ref{lemma:genfid}, we have that
\begin{equation}\nonumber
\begin{split}
  \F(\rho^{AB},m)&\ge\int\d g\ \frac{\Tr
    \omega^{A'E}_{m,g}+\Tr\omega^E_{m,g}}2 -\frac 12\int\d g\
  \N{\omega^{A'E}_{m,g}-\tau^{A'}_m\otimes\omega_{m,g}^E}_1\\
  &=1-\frac 12\int\d g\
  \N{\omega^{A'E}_{m,g}-\tau^{A'}_m\otimes\omega_{m,g}^E}_1,
\end{split}
\end{equation}
where, in the second line, we used the fact that $\int\Tr
\omega^{A'E}_{m,g}\d g=\int\Tr
\omega^{E}_{m,g}\d g=1$.

Now, for any fixed $\delta\ge0$, let
$\bar\rho^{AE}\in\B(\rho^{AE};\delta)$, where
$\rho^{AE}=\Tr_B[\Omega^{ABE}]$. Let us, moreover, define
$\bar\omega^{A'E}_{m,g}:=\frac {d_A}m(P_m^AU_g^A\otimes\openone_E)
\bar\rho^{AE} (P_m^AU_g^A\otimes\openone_E)^\dag$. By the triangle
inequality, we have that
\begin{eqnarray}
  \phantom{\le}&\N{\omega^{A'E}_{m,g}-\tau^{A'}_m\otimes\omega^E_{m,g}}_1&\nonumber\\
  \le&\N{\bar\omega^{A'E}_{m,g}-\tau^{A'}_m\otimes\bar\omega^E_{m,g}}_1&+\N{\omega^{A'E}_{m,g}-\bar\omega^{A'E}_{m,g}}_1\nonumber\\
  &&+\N{\tau^{A'}_m\otimes\bar\omega^E_{m,g}-\tau^{A'}_m\otimes\omega^E_{m,g}}_1\nonumber\\
  \le&\N{\bar\omega^{A'E}_{m,g}-\tau^{A'}_m\otimes\bar\omega^E_{m,g}}_1&+\N{\omega^{A'E}_{m,g}-\bar\omega^{A'E}_{m,g}}_1+\N{\bar\omega^E_{m,g}-\omega^E_{m,g}}_1.\nonumber
\end{eqnarray}
Since $\N{\bar\omega^E_{m,g}-\omega^E_{m,g}}_1\le
\N{\bar\omega^{AE}_{m,g}-\omega^{AE}_{m,g}}_1$, we have that
\begin{equation}\nonumber
\F(\rho^{AB},m)
\ge 1-\int\d g\N{\bar\omega^{A'E}_{m,g}-\tau^{A'}_m\otimes\bar\omega^E_{m,g}}_1-2\int\d g\N{\omega^{A'E}_{m,g}-\bar\omega^{A'E}_{m,g}}_1,
\end{equation}
for any choice of $\bar\rho^{AE}$ in $\B(\rho^{AE};\delta)$. Now,
thanks to Lemma~3.2 of Ref.~\cite{dec-capacity} and
eq.~\reff{fidelity}, we know that
\begin{equation}\nonumber
  \int\d g\N{\omega^{A'E}_{m,g}-\bar\omega^{A'E}_{m,g}}_1\le \N{\bar\rho^{AE}-\rho^{AE}}_1\le2\delta,
\end{equation}
which leads us to the estimate
\begin{equation}\nonumber
  \F(\rho^{AB},m)\ge 1-4\delta-\int\d g\N{\bar\omega^{A'E}_{m,g}-\tau^{A'}_m\otimes\bar\omega^E_{m,g}}_1.
\end{equation}
We are hence left with estimating the last group average. 

In order to do so, we exploit a technique used by Renner~\cite{thesis}
and Berta~\cite{mario}: by applying Lemma~\ref{lemma:tr-hs}, for any
given state $\sigma^E$ invertible on $\supp\ \bar\rho^E$, we obtain
the estimate
\begin{equation}\nonumber
\begin{split}
  \N{\bar\omega^{A'E}_{m,g}-\tau^{A'}_m\otimes\bar\omega^E_{m,g}}_1^2&\le m\Tr\left[(\bar\omega^{A'E}_{m,g}-\tau^{A'}_m\otimes\bar\omega_{m,g}^E)\ X_{m,g}^{A'E}\right]\\
  &:=m\N{\tilde{\rho}^{A'E}_{m,g}-\tau^{A'}_m\otimes\tilde{\rho}^E_{m,g}}^2_2,
\end{split}
\end{equation}
where 
\begin{enumerate}
\item $X_{m,g}^{A'E}:=
  (P^{A'}_m\otimes\sigma^E)^{-1/2}(\bar\omega^{A'E}_{m,g}-\tau^{A'}_m\otimes\bar\omega_{m,g}^E)(P^{A'}_m\otimes\sigma^E)^{-1/2}$,
\item $P^{A'}_m=P^A_m(P^A_m)^\dag=m\tau_m^{A'}$, see Eq.~(\ref{Pa}),
\item $\N{O}_2:=\sqrt{\Tr[O^\dag O]}$ denotes the Hilbert-Schmidt
  norm,
\item
  $\tilde\rho^{A'E}_{m,g}:=(P^{A'}_m\otimes\sigma^E)^{-1/4}\bar\omega^{A'E}_{m,g}(P^{A'}_m\otimes\sigma^E)^{-1/4}$,
  and finally,
\item
  $\tilde\rho^E_{m,g}:=\Tr_{A'}[\tilde\rho^{A'E}_{m,g}]=(\sigma^E)^{-1/4}\bar\omega^E_{m,g}
  (\sigma^E)^{-1/4}$.
\end{enumerate}
It is easy to check that
\begin{equation}\nonumber
  \N{\tilde\rho^{A'E}_{m,g}-\tau^{A'}_m\otimes\tilde\rho^E_{m,g}}^2_2=\N{\tilde\rho^{A'E}_{m,g}}_2^2-\frac 1m\N{\tilde\rho^E_{m,g}}_2^2.
\end{equation}
Further, using the concavity of the function $f(x) = \sqrt{x}$, we have
\begin{equation}\label{eq:average1}
\F(\rho^{AB},m)\ge 1-4\delta -\sqrt{\left\{m\int\d g\ \N{\tilde\rho^{A'E}_{m,g}}_2^2-\int\d g\ \N{\tilde\rho^E_{m,g}}_2^2
\right\}}.
\end{equation}
Standard calculations, similar to those reported
in~\cite{state-merg,dec-capacity,mario}, lead to
\begin{equation}\nonumber
\int\d g\ \N{\tilde\rho^{A'E}_{m,g}}_2^2=\frac {d_A}m\frac{d_A-m}{d_A^2-1}\N{\tilde\rho^{E}}_2^2+\frac {d_A}m\frac{md_A-1}{d_A^2-1}\N{\tilde\rho^{AE}}_2^2
\end{equation}
and
\begin{equation}\nonumber
\int\d g\ \N{\tilde\rho^{E}_{m,g}}_2^2=\frac {d_A}m\frac{m{d_A}-1}{{d_A}^2-1}\N{\tilde\rho^E}_2^2+\frac {d_A}m\frac{{d_A}-m}{{d_A}^2-1}\N{\tilde\rho^{AE}}_2^2,
\end{equation}
where
\begin{equation}\nonumber
\tilde\rho^{AE}:=(\openone_A\otimes\sigma^E)^{-1/4}\bar\rho^{AE}(\openone_A\otimes\sigma^E)^{-1/4},
\end{equation}
and $\tilde\rho^E:=\Tr_A[\tilde\rho^{AE}]$. By simple
manipulations, we arrive at
\begin{equation}\nonumber
 m\int\d g\ \N{\tilde\rho^{A'E}_{m,g}}_2^2-\int\d g\ \N{\tilde\rho^E_{m,g}}_2^2=\frac{{d_A}^2(m^2-1)}{m({d_A}^2-1)}\left\{\N{\tilde\rho^{AE}}_2^2-\frac 1{d_A}\N{\tilde\rho^E}_2^2\right\}.
\end{equation}
Since $m\le {d_A}$,
\begin{equation}\nonumber
  \frac{{d_A}^2(m^2-1)}{m({d_A}^2-1)}=m\frac{1-\frac{1}{m^2}}{1-\frac{1}{{d_A}^2}}\le m,
\end{equation}
so that eq.~(\ref{eq:average1}) can be rewritten as
\begin{equation}\nonumber
  \F(\rho^{AB},m)\ge 1-4\delta-\sqrt{
    m\left\{\N{\tilde\rho^{RE}}_2^2-\frac 1{d_A}\N{\tilde\rho^E}_2^2\right\}
  },
\end{equation}
for any choice of the states $\bar\rho^{AE}\in
\B(\rho^{AE};\delta)$ and $\sigma^E$ invertible on $\supp\
\bar\rho^E$.

Now, notice that
\begin{equation}\nonumber
  \N{\tilde\rho^{AE}}^2_2\le2^{S_2(\bar\omega^{AE}\|\openone_A\otimes\sigma^E)}.
\end{equation}
This inequality easily follows from~\reff{orig}, i.e.,
\begin{equation}\nonumber
\begin{split}
  \Tr[(\omega^{-1/4}\rho\omega^{-1/4})^2]&=\Tr[\omega^{-1/2}\rho\omega^{-1/2}\rho]\\
  &\le\Tr[\rho^2\omega^{-1}]=2^{S_2(\rho\|\omega)}.
\end{split}
\end{equation}
Moreover, from Lemma~\ref{lemma:tr-hs}, $\N{\tilde\rho^E}_2^2\ge
1$. Thus,
\begin{equation}\nonumber
  \F(\rho^{AB},m)\ge 1-4\delta-\sqrt{m\left\{2^{S_2(\bar\rho^{AE}\|\openone_A\otimes\sigma^E)}-\frac1{d_A}\right\}},
\end{equation}
for any choice of states $\bar\rho^{AE}\in \B(\rho^{AE};\delta)$ and
$\sigma^E$, the latter strictly positive on $\supp\bar\rho^{E}$. In
order to tighten the bound, we first optimize (i.e. minimize)
$S_2(\bar\rho^{AE}\|\openone_A\otimes\sigma^E)$ over $\sigma^E$ for
any $\bar\rho^{AE}$, obtaining $I^{A\to E}_2(\bar\rho^{AE}|E)$. We
further optimize (i.e. minimize) $I^{A\to E}_2(\bar\rho^{AE})$ over
$\bar\rho^{AE}\in\B(\rho^{AE};\delta)$, eventually obtaining
$I^{A\to E}_{2,\delta}(\rho^{AE})$.\end{proof}

\section*{Appendix B: Lemma~\ref{i2i0}}

\begin{lemma}\label{i2i0}
  For any pure state $\Psi^{ABE}$ of a tripartite system $ABE$, for
  any $\delta >0$, we have that \be\label{eyes} -I^{A\to
    E}_{2,\delta}(\rho^{AE})\ge I^{A\to B}_{0,\delta}(\rho^{AB}), \ee
  where $\rho^{AE}$ and $\rho^{AB}$ denote the corresponding reduced
  states of the subsystems $AE$ and $AB$, respectively; the smoothed
  $2$-coherent information $I^{A\to E}_{2,\delta}(\rho^{AE})$ is
  defined through \reff{coh-smooth}, and $I^{A\to
    B}_{0,\delta}(\rho^{AB})$ is the smoothed zero-coherent
  information given by~\reff{eq:i}.
\end{lemma}

\begin{proof}
  We make use of the fact that for any $P,Q\ge 0$,
\begin{equation}
  D_{\max}(P||Q) \ge S_2(P||Q),
\end{equation}
where $S_2(P||Q)$ is the relative-R\'enyi entropy of order $2$, and
$D_{\max}(P||Q)$ is the max-relative entropy between $P$ and $Q$
defined as follows \cite{nila}:
\begin{equation}
  D_{\max}(P||Q):=\log\min\{\lambda:P\le\lambda Q\}.
\label{smax}
\end{equation}
For any $\delta \ge0$, the smoothed conditional min-entropy is defined
as
\begin{equation}
H_{\min}^\delta(\rho^{AE}|E) := \max_{\bar\rho^{AE}\in
  \B(\rho^{AE};\delta)}\max_{\sigma^E \in \states(\sH_E)}\Bigl[ -
D_{\max}(\bar\rho^{AE}\|\openone^A \otimes \sigma^E) \Bigr]
\end{equation}
Moreover,
\begin{equation}
\label{min-min}
\begin{split}
&H_{\min}^\delta(\rho^{AE}|\rho^E)\\
:=&\max_{\bar\rho^{AE}\in \B(\rho^{AE};\delta)} H_{\min}(\bar\rho^{AE}
|\bar\rho^{E})\\
=&\max_{\bar\rho^{AE}\in \B(\rho^{AE};\delta)}\Bigl[ -
D_{\max}(\bar\rho^{AE}\|\openone^A \otimes \bar\rho^E) \Bigr]\\
=&\max_{\bar\rho^{AE}\in \B(\rho^{AE};\delta)}\left[-
D_{\max}\left(\bar\rho^{AE}\left\|\openone^A \otimes
    \frac{\bar\rho^E}{\Tr\bar\rho^E}\right.\right)+\log\Tr\bar\rho^E
\right]\\
\le& \max_{\bar\rho^{AE}\in \B(\rho^{AE};\delta)}\left[-
D_{\max}\left(\bar\rho^{AE}\left\|\openone^A \otimes
    \frac{\bar\rho^E}{\Tr\bar\rho^E}\right.\right)\right]\\
\le& H_{\min}^\delta(\rho^{AE}|E),
\end{split}
\end{equation}
where in the third equality we used the fact that
$D_{\max}(P\|Q)=D_{\max}(P\|cQ)+\log c$, for any $c\in\mathbb{R}$,
and, in the subsequent inequality, the fact that $\Tr\bar\rho^E\le
1$.

We also need the following duality relation, which was proved
in~\cite{mario} for two reduced states $\rho^{AB}$ and $\rho^{AE}$ of
the same tripartite pure state $\Psi^{ABE}$, but which can be extended
to sub-normalized states $\bar\rho^{AB}$ and $\bar\rho^{AE}$ coming
from $\bar\Psi^{ABE}$ as well:
\begin{equation} \label{mario}
  \begin{split}
    H_{\min}(\bar\rho^{AE}|\bar\rho^E) :&=-
    D_{\max}(\bar\rho^{AE}\|\openone_A \otimes \bar\rho^E)
    \\
    &= H_0(\bar\rho^{AB}|B),
\end{split}
\end{equation}
where $H_0(\bar\rho^{AB}|B):=-\min_{\sigma^B \in
  \states(\sH_B)}S_0(\bar\rho^{AB}\|\openone_A\otimes\sigma^B)$.

Further (as in the proof of Lemma 3 of~\cite{roger}), for any
$\delta\ge0$, let $ \B_*(\rho;\delta)$ denote the set of pure states
close to a state $\rho$, i.e.,
\begin{equation}
  \B_*(\rho;\delta) := \{\psi\in
  \B(\rho;\delta):\rank\psi=1\},
\end{equation}
and let
\begin{equation}
  \bar\B(\rho^{AB};\delta):= \left\{\tr_E(\bar\varphi^{ABE}) :\bar\varphi^{ABE} \in
    \B_*(\Psi^{ABE};\delta) \right\}
\end{equation}
where $\Psi^{ABE}$ is any arbitrarily fixed purification of
$\rho^{AB}$. Hence, $\bar\B(\rho^{AB};\delta)$ is the set of states
which are $\delta$-close to $\rho^{AB}$ (with respect to the fidelity)
on the purified space.  It was proved in~\cite{roger} that
\begin{equation}
\label{equal} \bar\B(\rho^{AB};\delta) =
\B(\rho^{AB};\delta).
\end{equation}
This is because on one hand the monotonicity of the fidelity under
partial trace ensures that $\bar\B(\rho^{AB};\delta) \subseteq
\B(\rho^{AB};\delta)$.  On the other hand , by Uhlmann's theorem
\cite{uhlmann}, every $\bar\rho^{AB} \in \B(\rho^{AB};\delta)$ has a
purification $\bar\varphi^{ABE} \in \B_*(\rho^{ABE};\delta)$, and this
implies that $\B(\rho^{AB};\delta)\subseteq\bar\B(\rho^{AB};\delta)$.

We now proceed to prove Lemma~\ref{i2i0}. From the
definitions~\reff{cond-smooth} and~\reff{coh-smooth} of $I^{A\to
  E}_{2,\delta}(\rho^{AE})$ and the $\alpha$-conditional entropies,
respectively, we have that
\begin{equation}
\begin{split}
  - I^{A\to E}_{2,\delta}(\rho^{AE}) &\equiv H_2^\delta(\rho^{AE}|E)\\
  &=\max_{\bar\rho^{AE}\in \B(\rho^{AE};\delta)}\max_{\sigma^E \in
    \states(\sH_E)}\Bigl[ - S_2(\bar\rho^{AE}\|\openone^A \otimes
  \sigma^E)
  \Bigr]\\
  &\ge\max_{\bar\rho^{AE}\in \B(\rho^{AE};\delta)}\max_{\sigma^E \in
    \states(\sH_B)}\Bigl[ - D_{\max}(\bar\rho^{AE}\|\openone^A \otimes
  \sigma^E)
  \Bigr]\\
  &\equiv H_{\min}^\delta (\rho^{AE}|E)\\
&\ge H_{\min}^\delta (\rho^{AE}|\rho^E)\\
  &= \max_{\bar\varphi^{ABE}\in \B_*(\Psi^{ABE};\delta)}H_{\min}
  (\bar\rho^{AE}|\bar\rho^E)\\
  &= \max_{\bar\varphi^{ABE}\in \B_*(\Psi^{ABE};\delta)}\Bigl[-H_0
  (\bar\rho^{AB}|B)\Bigr]\\
  &=I^{A\to B}_{0,\delta}(\rho^{AB}),
\end{split}
\end{equation}
where the second inequality follows from~\reff{min-min}, while the
equality at the sixth line follows from the fact that
$\bar\B(\rho^{AE};\delta)=\B(\rho^{AE};\delta)$. The
subsequent identity follows from~\reff{mario}, while the last identity
follows from~\reff{equal} and the definition of the smoothed
$0$-coherent information~\reff{eq:i}.
\end{proof}

\end{document}